\documentclass{emulateapj}

\usepackage{graphicx}
\usepackage{float}
\usepackage{amsmath}
\usepackage{epsfig,floatflt}
\usepackage[usenames,dvipsnames]{xcolor}
\usepackage{multirow}
\usepackage{enumerate}
\usepackage{natbib}

\begin{document}

\title{Grid-based exploration of cosmological parameter space with Snake}

\author{K. Mikkelsen\altaffilmark{1}, S. K. N{\ae}ss\altaffilmark{1} and H. K. Eriksen\altaffilmark{1,2}}

\email{kristin.mikkelsen@astro.uio.no}

\altaffiltext{1}{Institute of Theoretical Astrophysics, University of
  Oslo, P.O.\ Box 1029 Blindern, N-0315 Oslo, Norway}

\altaffiltext{2}{Centre of Mathematics for Applications, University of
  Oslo, P.O.\ Box 1053 Blindern, N-0316 Oslo, Norway}

\begin{abstract}
  
  We present a fully parallelized grid-based parameter estimation
  algorithm for investigating multidimensional likelihoods called
  Snake, and apply it to cosmological parameter estimation. The basic
  idea is to map out the likelihood grid-cell by grid-cell according
  to decreasing likelihood, and stop when a certain threshold has been
  reached. This approach improves vastly on the ``curse of
  dimensionality'' problem plaguing standard grid-based parameter
  estimation simply by disregarding grid-cells with negligible
  likelihood. The main advantages of this method compared to standard
  Metropolis-Hastings MCMC methods include 1) trivial extraction of
  arbitrary conditional distributions; 2) direct access to Bayesian
  evidences; 3) better sampling of the tails of the distribution; and
  4) nearly perfect parallelization scaling. The main disadvantage is,
  as in the case of brute-force grid-based evaluation, a dependency on
  the number of parameters, $N_{\textrm{par}}$. One of the main goals
  of the present paper is to determine how large $N_{\textrm{par}}$
  can be, while still maintaining reasonable computational efficiency;
  we find that $N_{\textrm{par}} = 12$ is well within the capabilities
  of the method. The performance of the code is tested by comparing
  cosmological parameters estimated using Snake and the WMAP-7 data
  with those obtained using CosmoMC, the current standard code in the
  field. We find fully consistent results, with similar computational
  expenses, but shorter wall time due to the perfect parallelization
  scheme.
\end{abstract}


\section{Introduction}
\label{sec:introduction}

Cosmological models are described in terms of a modest number
of cosmological parameters that reflect the underlying physical
processes of the universe. These are today routinely measured by
experiments such as WMAP \citep{jarosik:2011}, Planck (2011) and the
Sloan Digital Sky Survey \citep{york:2000} through likelihood
techniques. 

The most popular parameter estimation algorithm in the cosmology
community to date is the CosmoMC package \citep{lewis:2002}, which
maps out the cosmological parameter space using a Metropolis-Hastings
Markov Chain Monte-Carlo (MCMC) sampler. The computational cost of
this method is almost exclusively determined by the external
evaluation of the likelihood, which typically takes a few seconds per
evalution; the expense of the internal book-keeping operation is
completely negligible compared to this. A complete analysis of current
data sets typically requires $\mathcal{O}(10^5)$ evaluations,
resulting in an overall computational cost of 100-10,000 CPU hours,
depending on the particular problem.

This process can be sped up in two fundamentally different ways,
namely either by reducing the cost per likelihood evaluation, or by
reducing the number of likelihood evaluations required, and both cases
have already been explored extensively in the literature. Examples of
the former include CMBFit \citep{sandvik:2004}, PICO
\citep{fendt:2007}, COSMONET \citep{auld:2007}, sparse grids
\citep{frommert:2010} and PkANN \citep{agarwal:2012}, all of which
essentially build up a library of known cosmological models given a
set of parameters, and interpolate within this library using some
statistical method. Examples of the latter include MultiNest
\citep{feroz:2009} and APS \citep{daniel:2012}, both of which reduce
the number of likelihood evalutions through more efficient sampling
algorithms than the Metropolis-Hastings sampler.

In this paper, we present an algorithm that falls in the last
category, aiming to reduce the total number of likelihood evaluations
rather than the cost per evaluation. The initial idea of this paper is
based on the following reasoning: If the problem under consideration
involved only a one-dimensional likelihood, the mapping algorithm of
choice would be obvious -- one would simply evaluate the likelihood
over a one-dimensional grid. The resulting function is both easier to
work with than a set of samples, as produced by a MCMC algorithm, and
more accurate. Furthermore, it generally requires fewer evaluations,
because whereas an MCMC approach builds up the shape of the
distribution by counting how many samples fall in a given parameter
range (``bin''), the direct approach only needs to evaluate the
likelihood in a given bin once. In other words, the MCMC approach
spends most of the time evaluating the same likelihood points over and
over again, which can give the direct evaluation approach a
computational edge.

The vast majority of two-dimensional likelihoods are also mapped by
grid methods rather than MCMC methods, while for three or four
dimensions, the preferred approach is not clear. However, for higher
dimensions, virtually all cases are so far handled by MCMC methods. At
this stage, the so-called ``curse of dimensionality'' becomes highly
relevant, as the number of likelihood evaluations depends
exponentially on the number of dimensions. For instance, computing 100
grid points in each of five dimensions requires $100^5$ evaluation,
which is generally far too many for most problems.

However, in this paper we point out that this is not necessarily
true. The point is simply that the vast majority of the
high-dimensionality volume typically has negligible likelihood, and
therefore does not need to be evaluated in the first place. The trick
is to figure out which grid cells are relevant and which are not. If
this can be done both efficiently and robustly, all the useful
properties of normal grids are retained, and computational cost is not
compromised. Further, by virtue of not being a Markov chain, the
algorithm parallelizes trivially, leading to shorter overall
computational wall time, which is often even more critical for a given
analysis problem than the total CPU time.

\section{The Snake algorithm}
\label{sec:algorith}

\subsection{Algorithm}

The Snake algorithm is very simple, and easily explained in terms of a
few basic steps. To do so succinctly, it is useful to first define
some terminology:
\paragraph{The grid} The Snake algorithm operates in a virtual grid defined
  by an origo, $\theta_0$ and a grid cell size, $\Delta\theta$, for
  each parameter. All points in parameter space are referred to and
  stored as an $N_{\textrm{par}}$ array on the form $\theta_0 +
  \mathbf{k}\Delta\theta $, where $\mathbf{k}$ is an integer vector
  describing the multidimensional bin number.
\paragraph{The surface} Each point in the grid can be assigned to
  one of three groups, namely external, internal and surface
  points. External points are those that have not yet been considered;
  internal points are those for which likelihoods have been computed
  for both the point itself and all its neighbors; surface points are
  points that have been considered, but have at least one unexplored
  neighbor.
\paragraph{The repository} Each considered parameter point is stored as an
  object in a data structure called the repository. This is a
  two-dimensional dynamic list in which each row defines one grid
  point, and contains $\mathbf{k}$, the likelihood value, the
  locations of its neighbors in the linked list, and a logical flag
  specifying whether the point is currently on the surface.

Given these definitions, the Snake algorithm may be summarized as follows:
\begin{enumerate}
\item \emph{Initialization:} Insert $\theta_0$ into the repository.
\item \emph{Evaluation:} Consider the surface point with the highest likelihood,
  $\mathbf{h}$, and randomly pick one of its unexplored neighbors,
  $\mathbf{h}+\Delta$. Evaluate
  $\mathcal{L}(\theta_{\mathbf{h}+\Delta})$.
\item \emph{Surface update:} If $\mathbf{h}$ has no more unexplored
  neighbors, set its surface flag to false. Insert $\mathbf{h}+\Delta$
  into the repository; if all its neghbors have already been explored
  set its surface flag to false.
\item \emph{Convergence check:} If
  $\log\mathcal{L}(\theta_{\textrm{best-fit}})-\log\mathcal{L}(\theta_i)$ is
  smaller than a predefined threshold for all surface points, $i$,
  then exit. If not, go to (2).
\end{enumerate}

This stepping procedure leads to two distinct phases. First there is a
burn-in period in which the solver performs a greedy
maximum-likelihood search. Then, once the maximum has been found, the
area around the peak is investigated such that the surface grows
outwards according to the underlying likelihood distribution, until
the threshold is reached. A larger threshold ensures that the tails
are investigated more closely, but also means that more evaluations
need to the performed which is computationally expensive, thus the
threshold should be kept low, though still making sure the edges are
properly investigated.

The likelihood evaluation is by far the most time consuming part in
cosmological evaluations. This implies that the Snake algorithm can be
very efficiently parallelized. In the current implementation we have
adopted a master--slave parallelization strategy, in which one
processor maintains the repository, and the remaining processors only
perform likelihood evaluations for parameters provided by the
master. This ensures both a simple implementation as well as close to
perfect speed-up; after only a few initial iterations there are always
more than enough available surface points to keep all processors
occupied. Moreover, the communication between the master and slave is
minimal, consisting only of a parameter multiplet and a likelihood
return value.

As should be clear from the above, Snake is algorithmically trivial;
this is nothing but an old-fashioned likelihood grid evaluation. The
only somewhat intricate part is to implement efficient book-keeping,
which is necessary in order to maintain computational efficiency as
the number of data elements, $V$, in the repository increases. For
this purpose, we implement dictionaries, based on the C++ standard map
template. These maps store the combination of two values, the key and
the mapped value, and enable access to the mapped value by using the
corresponding key in constant time, as opposed to $\mathcal{O}(V)$ for
unsorted lists or $\mathcal{O}(\log V)$ for sorted lists.

Two such maps are implemented for the master in Snake. The first is
for keeping track of which points in parameter space correspond to
which iteration, and is used to check if the neighbors of the current
point have already been visited, and if so returns their iteration
index. The second map keeps track of which likelihood value
corresponds to which iteration and is sorted according to descending
likelihood such that the first point on the list will always be that
with the highest likelihood, and therefore the index of the maximum
likelihood on the surface is just the mapped value at the top of the
map. When a point becomes an interior point the corresponding entries
are removed from the two maps in order to keep these as short as
possible and to avoid getting stuck at the overall maximum likelihood.

\begin{figure}[tbp]
  \mbox{\epsfig{figure=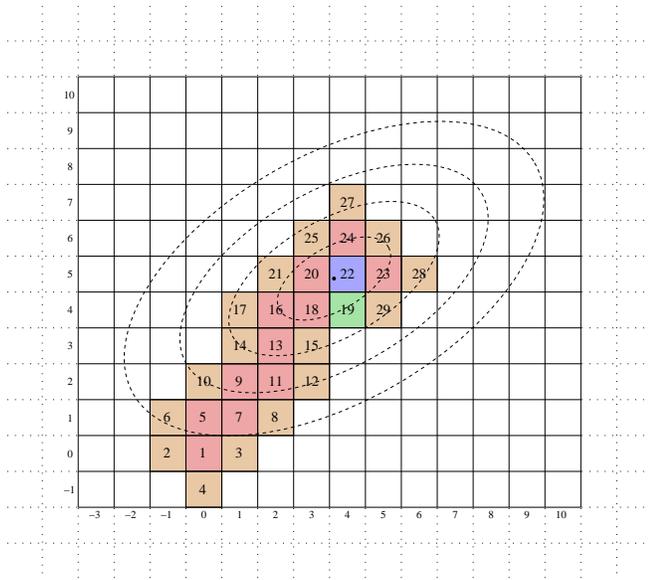,width=\linewidth,clip=}}
  \caption{A snapshot of a made up likelihood distribution after 29
    iterations illustrating the repository of Snake. Beige boxes are
    surface points, red are internal points, blue is the overall
    maximum likelihood, and green is the maximum likelihood on the
    surface.}
  \label{fig:array_illustration}
\end{figure}

\begin{deluxetable}{ccccccccc}
  \tablecaption{2D book repository \label{tab:arrays}}
  \tablecolumns{9} \tablehead{ \multirow{2}{*}{$i$} &
    \multicolumn{2}{c}{$\mathbf{k}$} & \multirow{2}{*}{$\mathcal{L}$} &
    \multicolumn{4}{c}{Ind} & \multirow{2}{*}{Surface} \\ \colhead{} &
    \colhead{x} & \colhead{y} & \colhead{} & \colhead{x:-1} &
    \colhead{x:+1} & \colhead{y:-1} & \colhead{y:+1} & \colhead{}}
  \startdata
  1 &  0 &  0 & 0.005 &  2 &  3 &  4 &  5 & F \\
  2 & -1 &  0 & 0.003 &    &  1 &    &  6 & T \\
  3 &  1 &  0 & 0.002 &  1 &    &    &  7 & T \\
  4 &  0 & -1 & 0.001 &    &    &    &  1 & T \\
  5 &  0 &  1 & 0.10  &  6 &  7 &  1 & 10 & F \\
  6 & -1 &  1 & 0.05  &    &  5 &  2 &    & T \\
  7 &  1 &  1 & 0.15  &  5 &  8 &  3 &  9 & F \\
  8 &  2 &  1 & 0.10  &  7 &    &    & 11 & T \\
  \hline \hline
  22 &  4 &  5 & 0.95  & 20 & 23 & 19 & 24 & F \\
  23 &  5 &  5 & 0.75  & 22 & 28 & 29 & 26 & F \\
  24 &  4 &  6 & 0.70  & 25 & 26 & 22 & 27 & F \\
  25 &  3 &  6 & 0.60  &    & 24 & 20 &    & T \\
  26 &  5 &  6 & 0.70  & 24 &    & 23 &    & T \\
  27 &  4 &  7 & 0.45  &    &    & 24 &    & T \\
  28 &  6 &  5 & 0.55  & 23 &    &    &    & T \\
  29 &  5 &  4 & 0.55  & 19 &    &    & 23 & T\enddata
\end{deluxetable}

\subsection{Walk-through of two-dimensional example}

Before testing the algorithm on realistic cases, it is useful to walk
through it step-by-step for a simple case, to gain some intuition for
its behaviour. In this section, we therefore first consider the small
two-dimensional example illustrated in Figure
\ref{fig:array_illustration} and Table \ref{tab:arrays}. The unknown
distribution to be mapped is marked in Figure
\ref{fig:array_illustration} by dashed lines, corresponding to 1, 2
and $3\sigma$ contours, and the threshold to be reached is defined as
the $3\sigma$ contour.

First, we initialize the code at $(0,0)$, which in this case happened
to lie slightly below and to the left of the maximum-likelihood
point. We evaluate the likelihood, and insert this point into the
first row of the repository (Table \ref{tab:arrays}). At this stage,
the first four columns are finalized, the surface flag is set to
\texttt{true}, and none of the neighbor indices (indicated by the
\texttt{ind} array of length $2N_{\textrm{par}}$) are set, indicating
that no neighbors have been evaluated yet.

Second, as specified by the algorithm we now find the surface point
with the highest likelihood, which of course is the point just
added. We select one of its neighbors, which in this case happened to
be $(-1,0)$. We evaluate its likelihood, and insert this new point
into the second row of the repository. We update the neighbor indices
of both this new point and the original point to point to each others
main index. We then repeat this process over and over again, adding
more and more points to the repository, until the smallest difference
between the likelihood of the overall maximum-likelihood point and
that of any point on the surface is larger than a pre-defined
threshold.

Table~\ref{tab:arrays} gives a snap-shot of the repository
(parameters, likelihood, current status of the \texttt{ind} array and
the surface flag) at iteration number 29, matching the illustration
seen in Figure \ref{fig:array_illustration}. The beige boxes
correspond to the points in parameter space which lie on the surface,
red boxes are interior points and the blue box corresponds to the
overall maximum likelihood. The green box is the parameter point on
the surface with the highest likelihood and will be the start point
for the next iteration. The numbers inside the boxes correspond to the
iteration index, thus the path Snake takes to reach the maximum
likelihood can be see, as well as the relation between neighbors and
the values of the first and last eight points quoted in the ind array
in Table~\ref{tab:arrays}. Iterations which have all \texttt{ind}
columns filled have their surface flag set to false and the point no
longer exists in the maps. The process continues until all boxes
touching the $3\sigma$ contour have turned red, after which the
surface lies fully below the threshold.

\subsection{Exploration of double-peaked likelihood}
\label{sec:2D}

A second illustration of how Snake investigates parameter space is
given by the double-peaked 2-dimensional likelihood 
\begin{equation}
  \mathcal{L} = A_1\textrm{e}^{\frac{1}{2} (\boldsymbol{x}-\boldsymbol{\mu_1})^T C_1^{-1} (\boldsymbol{x}-\boldsymbol{\mu_1})} + A_2\textrm{e}^{\frac{1}{2} (\boldsymbol{x}-\boldsymbol{\mu_2})^T C_2^{-1} (\boldsymbol{x}-\boldsymbol{\mu_2})},
\end{equation}
where $\boldsymbol{x}$ is the 2-dimensional parameter vector, $A_1$
and $A_2$ are the peak amplitudes, $C_1$ and $C_2$ the corresponding
covariance matrices and $\boldsymbol{\mu_1}$ and $\boldsymbol{\mu_2}$
the vectors of the means.

\begin{figure}[tbp]
  \mbox{\epsfig{figure=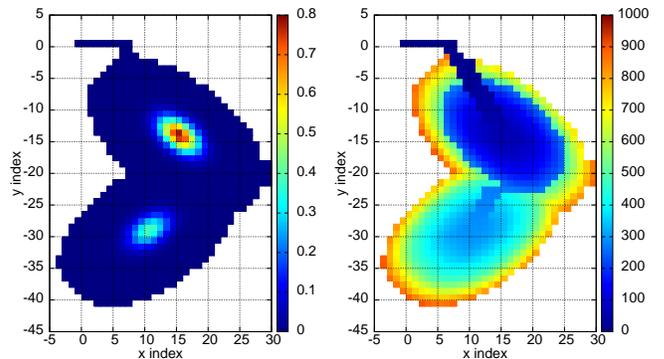,width=\linewidth,clip=}}
  \caption{Two-dimensional illustration of Snake's sampling
    method. \emph{Left}: The (unnormalized) target
    likelihood. \emph{Right}: The path Snake takes through parameter
    space. It finds the closest peak, investigates the area around
    this peak, discovers the second peak, investigates the area around
    this one, and finally explores the joint boundary of both peaks.}
  \label{fig:2D_toy}
\end{figure}

The leftmost plot of Figure~\ref{fig:2D_toy} shows a likelihood
distribution that can be described by this equation for a particular
set of covariance matrices and means. The path Snake takes in the
two-dimensional parameter space is shown in the rightmost plot of
Figure~\ref{fig:2D_toy} and as can be seen Snake quickly finds the
maximum likelihood of the closest peak, and then proceeds by
investigating the area around this peak by visiting neighbors of the
surface point with highest likelihood. When the likelihood being
investigated falls to the value corresponding to the intersection of
the two peaks Snake makes its way to the second peak, and continues by
investigating the area around this peak in the same manner as the
first peak. Once Snake returns to the likelihood equal to that at the
intersection it will investigate the points around both peaks until
the desired threshold is reached.

Note that if the two peaks had been so far apart that the likelihood
at the intersection fell below the threshold cutoff the second peak
would remain undiscovered. This problem can be solved in the same way
as for standard Metropolis-Hasting samplers: Run several Snakes in
parallel with different initial positions. Once two independent Snakes
touch for the first time, merge the repositories and the CPU working
groups into one master-slave organization.

\section{Accuracy and efficiency with increasing dimensionality}
\label{sec:gaussian}

The main outstanding question regarding the Snake algorithm is how
well it scales with the number of dimensions in terms of
efficiency. To study this question quantitatively, we consider a
correlated Gaussian likelihood on the form
\begin{equation}
  \mathcal{L} = \textrm{e}^{\frac{1}{2} (\boldsymbol{x}-\boldsymbol{\mu})^T C^{-1} (\boldsymbol{x}-\boldsymbol{\mu})},
\end{equation}
where $\boldsymbol{x}$, $C$ and $\boldsymbol{\mu}$ are the
multidimensional parameter vector, covariance matrix and vector of
means, respectively. Both the mean and standard deviation of dimension
number $i$ are arbitrarily chosen to be $i$.

\begin{deluxetable}{lrlrlc}
  \tablecaption{Cosmological parameters \label{tab:cosmological}}
  \tablecolumns{6}
  \tablehead{\colhead{Parameter} & \multicolumn{2}{c}{CosmoMC}
    & \multicolumn{2}{c}{Snake} & \colhead{Shift in $\sigma$}}
  \tablecomments{Comparison of best-fit parameters derived by CosmoMC
    and Snake from the 7-year WMAP data.}
  \startdata
  $\Omega_{b}h^2$ & 0.02252 & $^{+ 0.00055}_{- 0.00056}$ & 0.02252 & $^{+ 0.00057}_{- 0.00056}$ & 0 \\
  $\Omega_{DM}h^2$ & 0.1110 & $^{+ 0.0055}_{- 0.0054}$ & 0.1107 & $^{+0.0055}_{- 0.0054}$  & 0.06 \\
  $\theta$ & 1.039 & $\pm 0.003$ & 1.039 & $\pm 0.003$ & 0 \\
  $\tau$ & 0.08849 & $^{+ 0.00632}_{- 0.00754}$ & 0.08758 & $^{+0.01558}_{- 0.01426}$ & 0.08 \\
  $n_s$ & 0.9682 & $^{+ 0.0138}_{- 0.0136}$ & 0.9681 & $^{+ 0.0139}_{-0.0138}$ & 0.07 \\
  $\textrm{log}[10^{10} A_s]$ & 3.082 & $^{+ 0.034}_{- 0.035}$ & 3.080 & $\pm$ 0.035 & 0.06\enddata
\end{deluxetable}

Our goal is now to map out this distribution in $N_{\textrm{par}}$
dimensions, and determine the maximum number of dimensions that can be
probed with high accuracy using reasonable computational resources. To
do so, we impose a limit on the number of likelihood evaluations of $N
= 10^6$, a typical number for modern cosmological analyses. The grid
cell width in dimension $i$ is chosen to be $8i \times
N^{1/N_{\textrm{par}}}$, corresponding to distributing the $N$
evaluations roughly over a grid covering roughly $-4\sigma$ to
$+4\sigma$ in each of the $N_{\textrm{par}}$ dimensions. (Of course,
the actual shape probed by Snake will not be a rectangular grid, but
rather conform to the shape of the underlying distribution.) We then
run the algorithm for increasing $N_{\textrm{par}}$, and compare the
resulting marginals to the known analytic input marginals; once the
combined error in the derived mean or standard deviation is larger than
$0.1\sigma$, we consider the algorithm to have broken down as a result
of the sparse sampling of the underlying distribution.

\begin{figure}[tbp]
  \mbox{\epsfig{figure=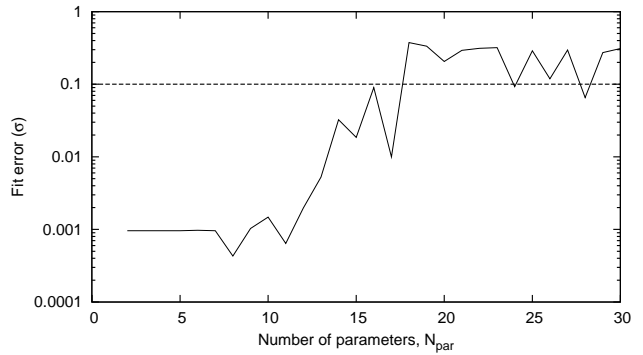,width=\linewidth,clip=}}
  \caption{Combined mean and standard deviation errors averaged over
    number of dimensions (solid) showing the $0.1\sigma$ cuttoff
    (dashed).}
  \label{fig:gaussian_limit}
\end{figure}

In Figure \ref{fig:gaussian_limit} we plot the combined mean and
standard deviation errors averaged over the number of dimensions,
$N_{\textrm{par}}$, as a function of $N_{\textrm{par}}$. Here we
clearly see that for $N_{\textrm{par}} < 12$, the algorithm recovers
the true distribution with high accuracy. Of course, given more
computational resources these errors can be decreased arbitrarily, but
since the cost faces an exponential growth with increasing
$N_{\textrm{par}}$, it seems reasonable to define the operational
range for Snake to be $N_{\textrm{par}} \leq 12-15$.

\section{7-year WMAP likelihood analysis}
\label{sec:cosmomogical}

\subsection{Parameter estimation}

\begin{figure}[tbp]
  \mbox{\epsfig{figure=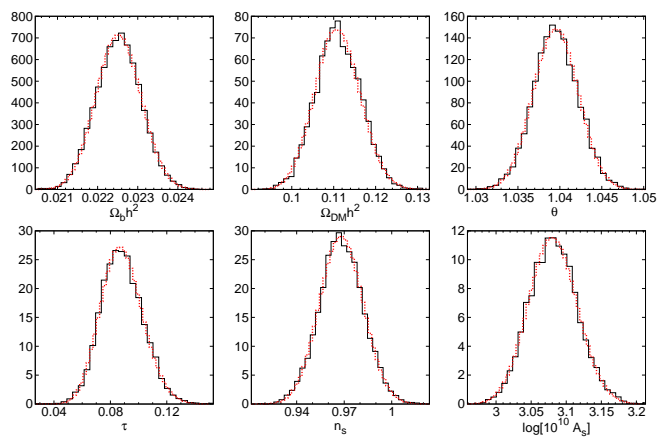,width=\linewidth,clip=}}
  \caption{Marginal cosmological parameter distributions derived with
    Snake (dashed red line) and CosmoMC (solid black line) from the
    7-year WMAP likelihood.}
  \label{fig:6params}
\end{figure}

We now apply this method to the 7-year WMAP likelihood, and estimate
cosmological parameters within the well-established 6-parameter
$\Lambda$CDM concordance model \citep{komatsu:2011}. The parameter set
of choice is $\Omega_{b}h^2$, $\Omega_{DM}h^2$, $\theta$, $\tau$,
$n_s$ and $\textrm{log}[10^{10}A_s]$. The same setup is analyzed using
both Snake and CosmoMC for comparison purposes.

The resulting normalized marginal distributions are shown in
Figure~\ref{fig:6params}, and means and standard deviations are
tabulated in Table~\ref{tab:cosmological}. The agreement between the
two methods is excellent, with a maximum difference between the two
methods corresponding to a $0.08 \sigma$ shift in $\tau$ and $0.07
\sigma$ shift in $n_s$.

The CosmoMC results were obtained with an MPI convergence criterion of
0.03, while the Snake convergence threshold was defined to be
-6.0. Both codes were run on 50 CPUs, and the resulting wall times
were 1.42 and 1.24 hours, respectively.

\subsection{Model selection by Bayesian evidence}
\label{sec:evidence}

A significant advantage of Snake over CosmoMC is its direct access to
the Bayesian evidence \citep[e.g.,][]{gelman:2003}. For a given model
$H$ with parameters $\theta$ and data $d$, this is simply the
normalization factor, $E \equiv P(d|H)$, in Bayes' theorem,
\begin{equation}
  \textrm{P}(\theta|d,H) = \frac{\textrm{P}(d|\theta,H)\textrm{P}(\theta|H)}{\textrm{P}(d|H)}.
\end{equation}
The other factors are the likelihood, $\mathcal{L}(\theta|H) =
\textrm{P}(d|\theta,H)$, the prior, $\textrm{P}(\theta|H)$, and the
posterior, $P(\theta|d, H)$. Different models can be compared in terms
of their evidence, which for a model, $H_n$, is given by
\begin{equation} \label{eq:evidence}
  \textrm{P}(d|H_n) = \int \limits_\Omega \textrm{P}(d,\theta|H_n) \textrm{d}\theta = \int \limits_\Omega \textrm{P}(d|\theta,H_n)\textrm{P}(\theta|H_n) \textrm{d}\theta,
\end{equation}
where $\textrm{P}(d,\theta|H_n)$ is the joint probability distribution
of $d$ and $\theta$ given this model over all of parameter space,
$\Omega$ with step sizes of $\textrm{d}\theta$.

Calculating the evidence for different models using results from Snake
is rather straightforward as the parameter space is gridded into even
cells of volume ${\int}d\theta$. The integral in equation
\ref{eq:evidence} becomes a sum of the likelihood values within the
threshold multiplied by the volume of one grid cell, where we assume a
uniform prior which gives a factor of $1/L$ for each parameter, where
$L$ is the range for each parameter.

To compare two different models, $H_1$ and $H_2$, it is common to
consider the quantity 
\begin{equation}
  \delta \textrm{log}E = \textrm{log}E_1 - \textrm{log}E_2
\end{equation}
where $E_1$ and $E_2$ are the evidences of models $H_1$ and $H_2$,
respectively. The larger the value of $\delta \textrm{log}E$ the
higher the evidence in favour of model $E_1$. To calibrate this
quantity, one commonly adopts the Jeffreys' scale
\citep{liddle:2006,trotta:2008},
\begin{tabular}{r r r l}
  & \multirow{3}{*}{\Bigg\{} & 1 & \quad evidence for $E_1$ is substantial \\
  $\delta \textrm{log}E >$ & & 2.5 & \quad evidence for $E_1$ is strong \\
  & & 5 & \quad evidence for $E_1$ is decisive.
\end{tabular}
\newline However, one should note that this scale only provides a
general guideline, and conclusions can be application specific; see,
e.g., \citet{nesseris:2012} for a recent discussion of this issue.

We now evaluate the evidence for both the standard six parameter model
described above and for a reduced model obtained by enforcing $n_s =
1$. We find that the individual evidences are $E_1 = -3743.16$ and
$E_2 = -3744.56$, respectively, with an estimated uncertainty in each
of 0.1. This corresponds to $\Delta \textrm{log}E$ of $1.40$ in favour
of the 6-parameter model; the full model therefore provides a better
fit to the data, even when accounting for the larger parameter
volume. Similar results have already been published by
\cite{parkinson:2010}. Note that given the full multi-dimensional
Snake likelihood, evaluation the evidence of all nested models is
trivial by similar calculations.

\section{Summary and outlook}
\label{conclusion}

In this paper we have described a simple grid-based estimator for
multi-dimensional likelihoods. This algorithm exploits the fact that
by far most of the $N_{\textrm{par}}$-dimensional parameter volume in
a general likelihood has negligible contributions, and spends its
computational resources only where the likelihood itself is
significant. However, in contrast to standard MCMC methods, it only
considers each parameter point once, relying on the actual value of
the likelihood.

The main advantages of this method are 1) trivial extraction of
arbitrary conditional distributions; 2) direct access to Bayesian
evidences; 3) better sampling of the tails of the distribution; and 4)
nearly perfect parallelization scaling. The main disadvantage is a
computational cost increasing exponentially with
$N_{\textrm{par}}$. However, we have shown that the algorithm is fully
capable of probing at least $N_{\textrm{par}} \lesssim 12-15$ with
reasonable computational resources, which is sufficient for current
cosmological models.

In the current implementation the total cost of the method is
comparable to that of CosmoMC for similar convergence
criteria. However, the cost for a full Snake analysis can be vastly
reduced by introducing adaptive grids, in which the grid cell depends
on the local properties of the likelihood, such that high-significance
regions are sampled more densely than the tail regions. The results
from this extension will be reported in a future publication.

\begin{acknowledgements}
The computations presented in this paper were carried out on Titan,
a cluster owned and maintained by the University of Oslo and
NOTUR. HKE acknowlegdes support from the ERC Starting Grant 
StG2010-257080. 
\end{acknowledgements}


\begin{thebibliography}{}

\bibitem[Agarwal et al.(2012)]{agarwal:2012} Agarwal, S., Abdalla,
  F.~B., Feldman, H.~A., Lahav, O., \& Thomas, S.~A.\ 2012, \mnras,
  424, 1409

\bibitem[Auld et al.(2007)]{auld:2007} Auld, T., Bridges, M., Hobson,
  M.~P., \& Gull, S.~F.\ 2007, \mnras, 376, L11

\bibitem[Daniel et al.(2012)]{daniel:2012} Daniel, S.~F., Connolly,
  A.~J \& Schneider, J.\ 2012, ArXiv e-prints, 1205.2708
  
\bibitem[Fendt \& Wandelt(2007)]{fendt:2007} Fendt, W.~A., \& Wandelt,
  B.~D.\ 2007, \apj, 654, 2

\bibitem[Feroz et al.(2009)]{feroz:2009} Feroz, F., Hobson, M.~P. \&
  Bridges, M.\ 2009, \mnras, 398, 1601

\bibitem[Frommert et al.(2010)]{frommert:2010} Frommert, M.,
  Pfl{\"u}ger, D., Riller, T., et al.\ 2010, \mnras, 406, 1177

\bibitem[Gelman et al.(2003)]{gelman:2003} Gelman, A., Carlin, J.~B.,
  Stern, H.~S. \& Rubin, D.~B.\ 2003, Dayesian Data Analysis (2nd
  ed.; Champan \& Hall/CRC)

\bibitem[Jarosik et al.(2011)]{jarosik:2011} Jarosik, N. et al.\ 2011,
  \apjs, 192, 14

\bibitem[Komatsu et el.(2011)]{komatsu:2011} Komatsu, E., Smith,
  K.~M., Dunkley, J., et al.\ 2011, \apjs, 192, 18

\bibitem[Lewis \& Bridle(2002)]{lewis:2002} Lewis, A., \& Bridle, S.\
  2002, \prd, 66, 103511

\bibitem[Liddle et al.(2006)]{liddle:2006} Liddle, A., Mukherjee,
  P. \& Parkinson, D.\ 2006, Astronomy and Geophysics, 47, 4, 040000-4

\bibitem[Nesseris \& Garcia-Bellido(2012)]{nesseris:2012} Nesseris,
  S., \& Garcia-Bellido, J.\ 2012, arXiv:1210.7652

\bibitem[Parkinson \& Libble(2012)]{parkinson:2010} Parkinson, D. \&
  Liddle, A.~R.\ 2010, \prd, 82, 10, 103533

\bibitem[Planck(2011)]{planck:2011} Planck Collaboration\ 2011, A\&A,
  536, 1

\bibitem[Sandvik et al.(2004)]{sandvik:2004} Sandvik, H.~B., Tegmark,
  M., Wang, X., \& Zaldarriaga, M.\ 2004, \prd, 69, 063005

\bibitem[Trotta(2008)]{trotta:2008} Trotta, R.\ 2008, Contemporary
  Physics, 49, 71-104

\bibitem[York et al.(2000)]{york:2000} York, D. G., Adelman, J.,
  Anderson, Jr., J. E., et al. 2000, AJ, 120, 1579
\end{thebibliography}
\end{document}